\documentclass[9pt,twocolumn]{extarticle}
\pdfoutput=1

\usepackage{soul}
\usepackage{cite}

\usepackage{graphicx}
\usepackage[margin=1in]{geometry}
\usepackage[usenames,dvipsnames]{color}
\usepackage{amsmath,amssymb}
\usepackage{flushend}
\usepackage{lipsum}

\usepackage{textcomp, gensymb}

\usepackage{mathpazo} 
\usepackage{courier} 
\normalfont
\usepackage[T1]{fontenc}
\usepackage{caption}
\usepackage{upgreek}

\newcommand{\ad}[1]{\textsuperscript{#1}\kern-2pt}

\def\({\left(}
\def\){\right)}
\def\[{\left[}
\def\]{\right]}

\usepackage{capt-of}

\def\({\left(}
\def\){\right)}
\def\[{\left[}
\def\]{\right]}

\usepackage{setspace}

\def\mytitle{
$60$dB high-extinction auto-configured Mach--Zehnder interferometer
}

\title{\vspace{-1.0cm}\Huge\textbf{\textsf{\mytitle}}}

\author{C.~M.~Wilkes$^{1,*}$, X.~Qiang$^{1}$,  J.~Wang$^{1}$, R.~Santagati$^{1}$, S.~Paesani$^{1}$, X.~Zhou$^{2}$,\\ D.~A.~B. ~Miller$^{3}$, G.~D. Marshall$^{1}$, M.~G. Thompson$^{1}$, and J.~L.~O\textquoteright Brien$^{1,3,\dagger}$}

\date{}
\begin{document}
\twocolumn[{%
\maketitle
\begin{center}
\begin{minipage}{0.7\textwidth}
\begin{center}
\vspace{-5mm}
\textsf{\footnotesize\textsuperscript{1}Centre for Quantum Photonics, H. H. Wills Physics Laboratory and Department of Electrical and Electronic Engineering, University of Bristol, Merchant Venturers Building, Woodland Road, Bristol, BS8 1UB, UK. 
\\\textsuperscript{2} State Key Laboratory of Optoelectronic Materials and Technologies and School of Physics and Engineering, Sun Yat-sen University, Guangzhou 510275, China
\\\textsuperscript{3} Ginzton Laboratory, Spilker Building, Stanford University, 348 Via Pueblo Mall, Stanford, California 94305-4088, USA
\\
}

\end{center}
\end{minipage}

\end{center}

\setlength\parindent{12pt}

\begin{quotation}
\noindent
\textbf{Abstract:}
Imperfections in integrated photonics manufacturing have a detrimental effect on the maximal achievable visibility in interferometric architectures. These limits have profound implications for further photonics technological developments and in particular for quantum photonics technologies.
Active optimisation approaches, together with reconfigurable photonics, have been proposed as a solution to overcome this. 
In this paper, we demonstrate an ultra-high (>$60$ dB) extinction ratio in a silicon photonic device consisting of 
cascaded Mach-Zehnder interferometers, in which additional interferometers function as variable beamsplitters. 
The imperfections of fabricated beamsplitters are compensated using an automated progressive optimization algorithm with no requirement for pre-calibration. This work shows the possibility of integrating and accurately controlling linear-optical components for large-scale quantum information processing and other applications.

\vspace{0.2cm}

\end{quotation}
}]
  
\section*{Introduction} 
Photonics technology has been advancing strongly recently, especially with platforms like silicon photonics~\cite{Thomson:2016}, with many potential applications in classical interconnects, communications and sensing~\cite{Agrell:2016}, where it offers solutions to increasing problems of interconnect density and energy inside machines~\cite{Miller:2009,Krishnamoorthy:2015}, complements the processing abilities of electronics~\cite{ITRS:2015}, and offers novel possibilities like mode-division multiplexing~\cite{Agrell:2016} at long distances. Also, photonics is considered a promising physical implementation for quantum information technologies including quantum communication~\cite{bennett1984quantum,korzh2015provably}, metrology~\cite{holland1993interferometric,dowling2008quantum}, and computation~\cite{knill2001scheme,NielsenChuang,OBrienreview}. This is owing to the photon's properties of long coherence time and ease of manipulation~\cite{Politi,Shadbolt,Reck,Silverstone2014,SiBell,Sharping,Tang,Dirk,Wang2015,QKD}. A vast array of key components in integrated quantum photonics (IQP) have been reported in recent years including integrated photon sources~\cite{SiBell, Sharping,Silverstone2014}, linear optical circuits~\cite{Reck,Shadbolt}, and waveguide photon detectors~\cite{Tang,Dirk}. Many such uses of photonic technology, including sensing, mode conversion, IQP, and linear optical circuits generally, require quite precise interference of beams, for example in Mach-Zehnder interferometers (MZIs). A key difficulty is, however, that it is difficult to fabricate such components very precisely; a particular problem that limits performance, such as extinction ratios, is beamsplitter splitting ratio imperfections. Fortunately, the technique demonstrated in this paper compensates for such imperfections, making photonics devices tolerant to these faults.

The MZI, typically consisting of one phase-shifter with two outer beamsplitters, is one of the most crucial components in integrated photonics~\cite{Chen,Suzuki:2014}. It is used in several fields from metrology and wavelength multiplexing, through to quantum photonics ~\cite{Shadbolt,Reck, Silverstone2014,SiBell} and quantum communications~\cite{QKD,Wang2015}. As the integration density of photonic components increases, high-performance MZIs will become an extremely valuable resource because any operation error of a single MZI will propagate along the circuit and accumulate exponentially. This is especially true for fault-tolerant linear optical quantum computing\cite{Raussendorf,Terry,Terry2}. The quality of a high-performance MZI is typically quantitatively described by its extinction ratio. 
The extinction ratio, in the presence of indistinguishable photons, is a measure of how well one can distinguish between orthogonal computational basis states, and it is fundamentally determined by the splitting ratios of the beamsplitters used in the MZIs themselves~\cite{Shadbolt,Reck}. 
Because of fabrication imperfections, integrated beamsplitters usually have different splitting ratios to the desired ones---sometimes quite far away from 50:50~\cite{Chen,Suzuki:2014,Suzuki:2015}. A possible solution is presented by the combined use of active components together with automated optimisation  techniques~\cite{Miller:2013, Miller:2015}.  
In this way, it is possible to compensate for fabrication imperfections in order to achieve high-visibility interference in MZIs.

\begin{figure}
\includegraphics[scale = 0.63]{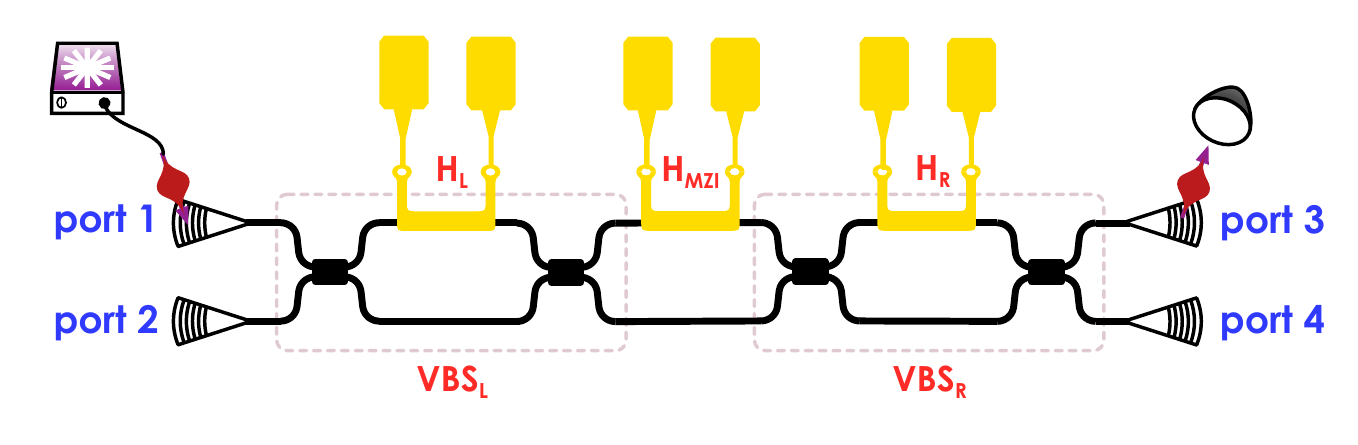}
\setlength{\abovecaptionskip}{-10pt}
\setlength{\belowcaptionskip}{-0pt}
\caption{
Self-optimized MZI and experimental setup. The cascaded MZI device was fabricated in silicon photonic waveguides. It consists of 3 thermo-optical phase-shifters and 4 multi-mode interference (MMI) beamsplitters. The $H_L$ and $H_R$ thermal phase-shifters, in conjunction with their adjacent beamsplitters, form the variable beamsplitters (VBS) (dashed boxes), and $H_{MZI}$ phase-shifter offers the usual control of the MZI interference. 
The progressive algorithm was performed on the $H_L$ and $H_R$  
to search their voltages targeting perfect $VBS_L$ and $VBS_R$. The central 
$H_{MZI}$ was scanned to verify the high extinction nature. 
}
\label{fig:ExpSetup}
\end{figure}
  
Here we present the first experimental demonstration of a near-perfect MZI on a silicon photonic chip by implementing the adaptive self-optimized approach proposed in Ref.\cite{Miller:2015}. This method consists of using two additional MZIs acting as reconfigurable variable beamsplitters (VBS) and finding their optimal configuration, thereby circumventing the requirements for perfect fabricated beamsplitters. This progressive approach allows automatic optimisation of the splitting ratios of two VBS to 50:50---without any prior calibration---and results in a high-extinction ratio MZI. Using a silicon photonic device, we measured $60.5$ dB extinction in the interferometric fringes, corresponding to an increase of $29.6$ dB with respect to the non-optimized case.
This work represents the new state of the art in the achievable extinction for interference, and a new approach for the realization of ultra-high fidelity operations for future fault-tolerant linear optical quantum computing~\cite{Terry2}.

\section*{Self-optimized Approach}

Figure \ref{fig:ExpSetup} shows the schematic of the MZI consisting of 7 components each with their own parameter: 3 thermo-optical phase-shifters ($H_L$, $H_R$ and $H_{MZI}$) with phases that we scan over, and 4 passive beamsplitters with fixed splitting ratios that are not of direct interest, as long as they fall within a range of 85:15 - 15:85~\cite{Miller:2015}. The phase-shifters are resistive heating elements and they induce a local change in the refractive index of the waveguide core by temperature variation. This allows us to optimize and control the MZI by applying electrical voltage onto each phase-shifter. To achieve an optimized MZI, we want to find the voltage settings for the outer phase-shifters, $H_L$ and $H_R$, that construct the VBS to be beamsplitters of 50:50 splitting ratio. 
Once this has been determined, the $H_L$ and $H_R$ can be set to the constant phase value and the central $H_{MZI}$ varied as a usual phase-shifter in a MZI. 

To find the optimum voltage settings for $H_L$ and $H_R$ we use an algorithm that enables us to set both $VBS_L$ and $VBS_R$ to 50:50 reflectivity, without pre-calibrating any component and only by simply minimizing or maximizing optical power in one output port~\cite{Miller:2015}. Considering the optical power output at port $3$, and rephrasing the reflectivities (i.e. splitting ratios) in their offset from the ideal 50:50 case ($\delta R_i = R_i - 0.5$),

\begin{equation} 
P^{\left(3 \right)} = \frac{1}{2} + 2 \left\lbrace \delta R_L \delta R_R - \left[\left(\frac{1}{4} - \delta R_L^2 \right)\left(\frac{1}{4} - \delta R_R^2 \right)\right]^{\frac{1}{2}}  \cos\theta \right\rbrace
\label{eq:transferfunction}
\end{equation}

\noindent where $R_i$ is the fabricated reflectivity of each VBS, $\theta$ is the phase of $H_{MZI}$, and $P^{\left(3 \right)}$ is the normalized measured optical power at port 3. The maximum and minimum power respectively correspond to $\theta = \pi$ and $\theta = 0$, with the cosine of $\theta$ therefore giving a sign change in the last term of Eq.~\eqref{eq:transferfunction}. It is at these settings of the $H_{MZI}$ phase that we determine the optimal voltages for $H_L$ and $H_R$.
The problem of configuring the VBS is thus reduced to a two-dimensional optimisation problem over the two VBS' reflectivities, $R_L$ and $R_R$. The convergence point of the algorithm~\cite{Miller:2015} is exactly located at the point where both reflectivities are 0.5 --- that is $\delta R_L$ and $\delta R_R$ are zero.

We use the algorithm of~\cite{Miller:2015}, though we have to extend it here because we do not have reliable prior knowledge of exactly what range of phase shifter voltages corresponds to monotonic increase or decrease of the splitting ratios of $VBS_L$ and $VBS_R$; for each such beamsplitter, that monotonic range will only occur over a specific range of size $\pi$ somewhere in a total $2 \pi$ range of phase shifts. So, we end up exploring the whole $2 \pi$ range for both beamsplitters to minimize or maximize power. We do this for the first pass of the algorithm; with subsequent passes we adapt the scan range to a smaller one centred on previously obtained voltage values. We start with all voltages set to zero. Hence our extended algorithm becomes:

\begin{enumerate}

\item Scan over the $2 \pi$ voltage range of
$H_{MZI}$ to obtain the voltage for the minimum power at output port $3$, $P_{min}^{\left(3\right)}$. 

\item Scan $H_L$ and $H_R$ over their voltage ranges in both equal and opposite directions to find minimum power: $\left( \delta V_{H_L}, \delta V_{H_R} \right) \ni \left\lbrace\left( +, + \right),  \left( -, - \right)\right\rbrace$, where $+$ means up, and $-$ means down.

\item Scan over the $2 \pi$ voltage range of
$H_{MZI}$ to obtain the voltage for the maximum power at output port $3$, $P_{max}^{\left(3 \right)}$.

\item Scan $H_L$ and $H_R$ over their voltage ranges in all 4 directions to find the maximum optical power: $\left( \delta V_{H_L}, \delta V_{H_R}\right) \ni \left\lbrace\left(+, +\right), \left(+, -\right), \left(-, +\right), \left(-, -\right)\right\rbrace$.

\item Repeat steps 1--4 until there are no further statistically significant and measurable changes in $V_{H_L}, V_{H_R}$ settings.\newline
\end{enumerate}

\noindent During the algorithm's operation all voltages are kept at their previously obtained values apart from the voltages modified in the current step.

Figure \ref{fig:ConvData} (a) depicts a typical calculated optical power response for a VBS MZI when it is subject to variation in the phase-shifter voltages, $V_L$ and $V_R$. The region plots were obtained by evaluating Eq. \ref{eq:transferfunction} for both the maximum and minimum power cases and equating them to the maximum (minimum) normalized power of 1 (0). The lines shown are contours for optical powers within a specified range of values. This threshold value is  0.995 for the maximum case, and 0.003 for the minimum case.

\section*{Experimental Results}
The MZI device was fabricated on a standard Silicon-on-Insulator wafer using 248 nm photo-lithography technology. 
The beamsplitters use a multi-mode interference (MMI) structure and thermo-optical phase-shifters were formed using TiN resistive heaters. The as-fabricated MMIs possess various splitting ratios that deviate from the ideal 50:50. 
The chip was mounted on a thermo-electrically controlled copper plate for stable temperature maintenance. Waste heat was dumped to a large copper heat sink. Each phase-shifter has a separate signal pin and mutually referenced ground pin, and was controlled by an individual voltage driver. Thermal crosstalk was negligible, and the device remained in a steady thermal state throughout.

Laser light at $1550.8$ nm wavelength from a Tunics T100S-HP laser was coupled to the chip (in at port $1$ and out at $3$) using grating couplers and a single-mode fiber array mounted on a piezo-controlled 6-axis translation stage. 
The grating coupler here also works as an on-chip transverse-electric (TE) polarizer~\cite{Taillaert:2003}, ensuring that all MZI components only interact with the TE-polarized light and thereby avoiding visibility degradation from an impure polarization state meeting in the MMIs. Before (after) light enters (leaves) the chip, it passes through a dense-wavelength-division multiplexer (DWDM) filter to reduce the amplified spontaneous emission noise from the laser and background signal from flat spectrum scattered light across the slab mode of the chip. Light is detected by a high-sensitivity Thorlabs S155C InGaAs powermeter with a dynamic range of $80$dB. All resistive heaters are controlled by a 12-bit in-house digital-to-analog converter (DAC) electrical driver, which enables us to optimize $H_L$ and $H_R$ and scan $H_{MZI}$ with high accuracy~\cite{QSim2016a}.

\begin{figure}[t!]
\includegraphics[scale = 0.25]{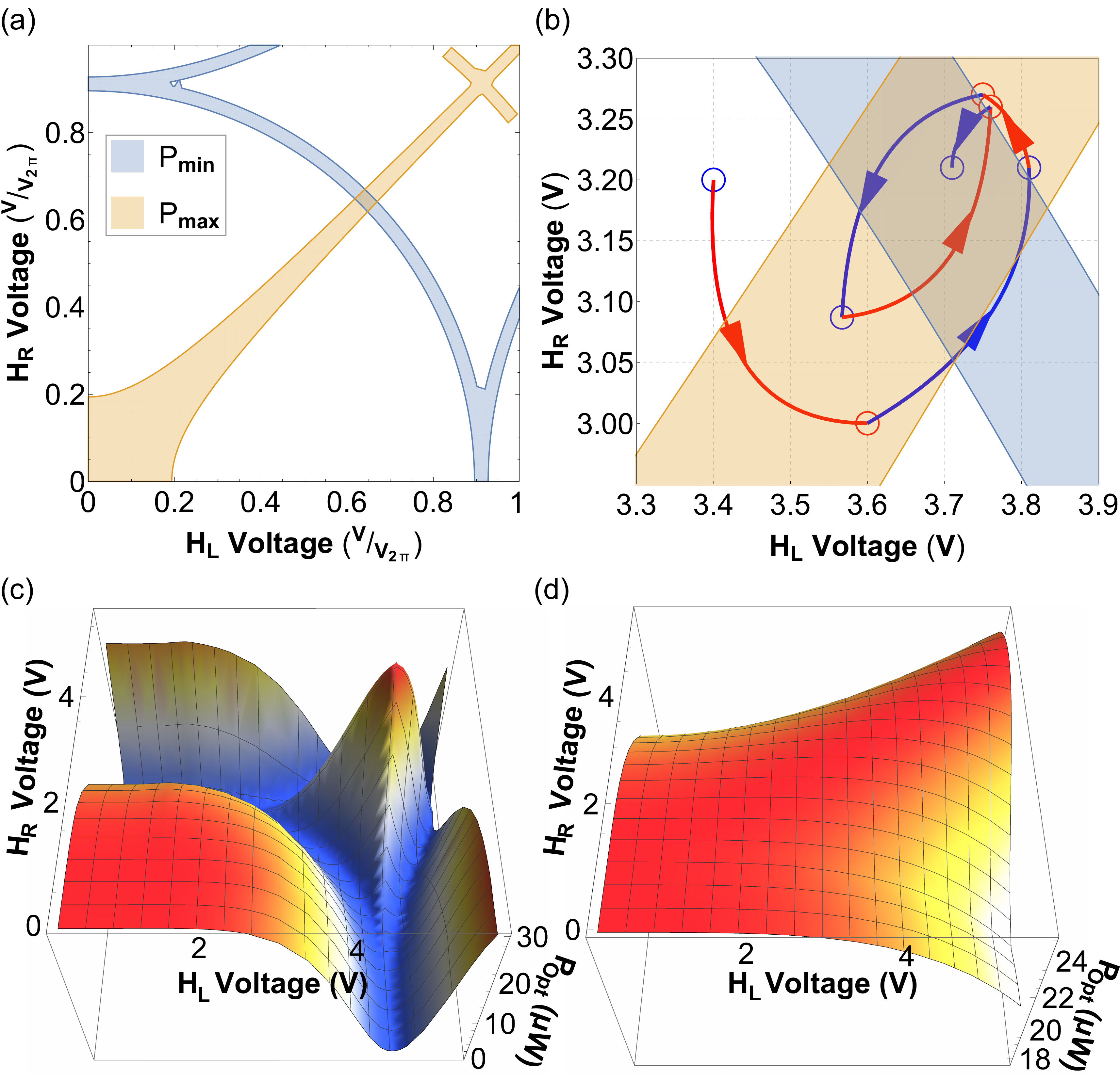}
\caption{
a) Simulation of the beamsplitter 50:50 setup algorithm. 
Arbitrary phase-shifter parameters were chosen for the algorithm simulation. 
Blue (orange) area represents the minimum (maximum) optical power $P_{min}$ ($P_{max}$) at port 3 with a change of $V_{H_L}$ and $V_{H_R}$.  
The algorithm ultimately converges to the intersection point of the blue and orange areas. 
Note that there are periodic solutions; 
the region shown is chosen within the first $2\pi$ phase voltage of both $H_L$ and $H_R$. This convergence point is easier to locate as phase scales with the square of the voltage, meaning that accurately achieving a particular phase becomes more difficult with increasing voltage.
b) Experimental convergence of the self-optimized algorithm.
Each point plotted indicates the ${H_L}$ and ${H_R}$ voltage settings obtained after steps 2 (blue) and 4 (red) for subsequent passes of the algorithm. (The colored arrow paths are guides for the eye, but the points are experimental results at the end of each stage of the algorithm.) Rapid convergence occurred, with the optimal point at $\left( 3.71 \textrm{ V, }3.21 \textrm{ V} \right)$, determining the 50:50 configuration of both $VBS_L$ and $VBS_R$. 
c) and d) The fitted surfaces from experiment data for minimum and maximum optical power, confirming the expected convergence lines shown in a).  
}
\label{fig:ConvData}
\end{figure}

The algorithm for finding the 50:50 VBS configuration was run on the device. The demonstration's results shown in Fig~\ref{fig:ConvData}(b) indicate a rapid convergence, caused by finding the region of the global extrema in the algorithm's first pass. This was achieved using the smallest voltage resolution available from the voltage driver ($5$ mV). This was limited by noise reducing the effective number of bits to 11 over a $10$ V range. Here, both $VBS$ were optimized as close to 50:50 allowable by both the algorithm and the equipment used. The subsequent MZI fringe measurement is a means of determining whether the beamsplitters are optimized or not. Throughout the experiment, light was injected in to port $1$, and the optical power was measured only at port $3$. This was important for demonstrating the capabilities of the algorithm, and that it only requires a single optical signal output to inform the voltage settings for subsequent steps.

Figs \ref{fig:ConvData} (c) and (d) are the 2D fits of experimental data over the $2\pi$ voltage space of the phase-shifters for the maximum and minimum optical power cases. These confirm the theoretical plot shown in Fig \ref{fig:ConvData} (a). The fits are obtained by considering a cubic term in the phase-voltage relationship of the phase-shifters, which has also been considered in characterisation of other integrated photonic devices~\cite{Shadbolt}. The cubic term encapsulates all non-Ohmic behaviour of the phase-shifters, including the change in their resistance when subject to higher temperatures.

\begin{figure}[t!]
\includegraphics[scale = 0.25]{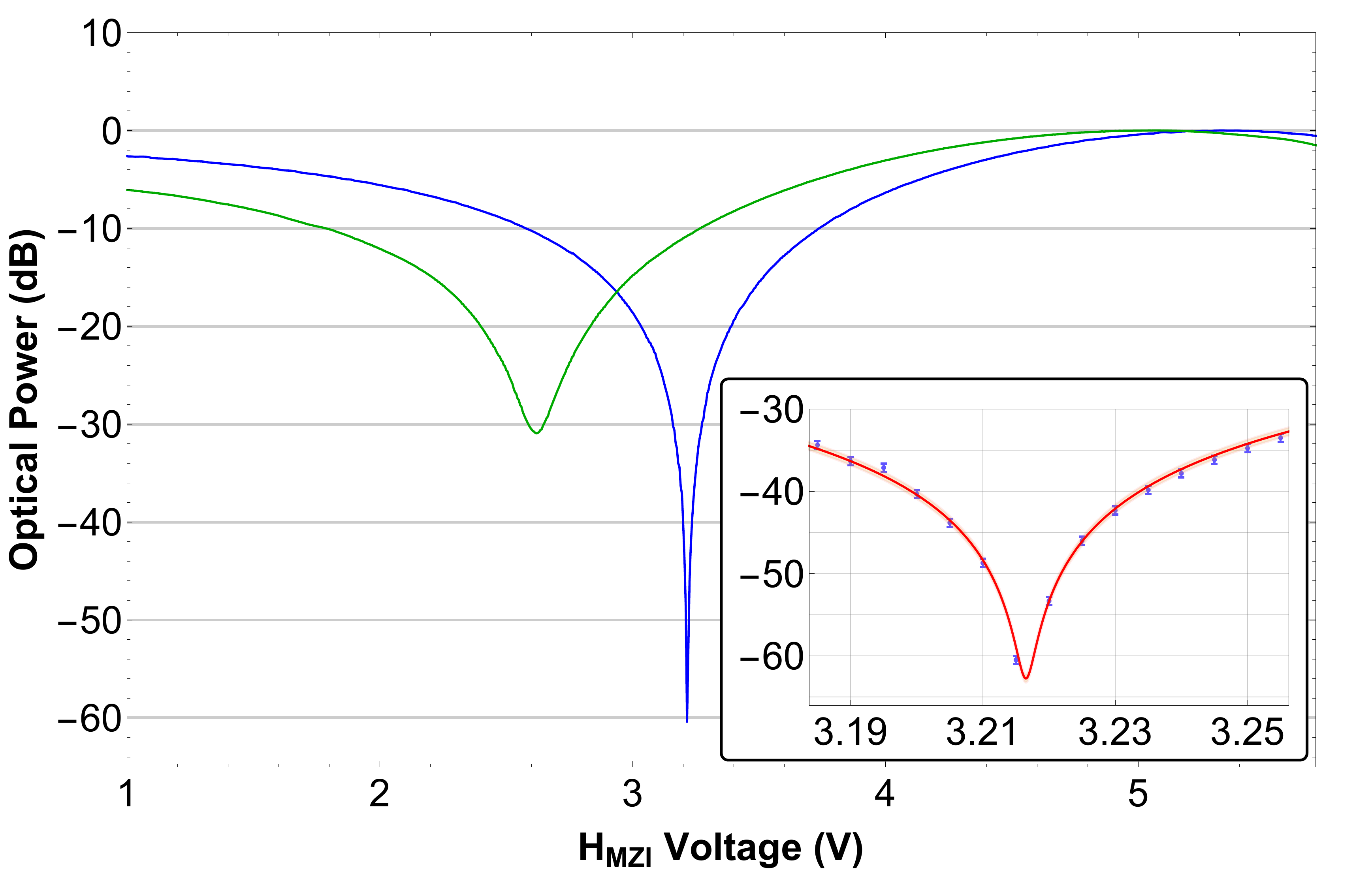}
\caption{
High extinction ratio MZI interference fringe. 
\emph{Blue}: Fringe obtained using self-optimized beamsplitters MZI, with $60.5$ dB extinction ratio. 
\emph{Green}: Fringe obtained using a single MZI with as-fabricated beamsplitters, with $30.9$ dB extinction ratio. Both fringes were taken under the same experimental condition. The enhancement in the extinction ratio by $29.6$ dBm indicates the significant improvement yielded by the self-optimising approach.      
\emph{Inset}: close-up of the fringe's dip, showing that the power readings in that region are well-behaved and are not due to any noise in the measurement apparatus.
All extinction ratio values used were obtained from the raw data values; no interpolated data from curve fitting is used.
}
\label{fig:HighXFringe}
\end{figure}

With the use of the self-optimized $VBS_L$ and $VBS_R$ we then implement the MZI. The MZI yielded an ultra-high extinction ratio of $60.5$ dB (Fig.~\ref{fig:HighXFringe}). For comparison, a fringe for a single MZI on the same chip with two fixed MMI beamsplitters was also obtained with a $30.9$ dB extinction ratio, which corresponds to beamsplitters with a 48.4:51.6 splitting ratio. This illustrates the significant improvement offered --- $29.6$ dB in this case --- by compensating for fabrication imperfections. Moreover, this demonstration here is a $10.1$ dB increase from the previous reported result using an active device consisting of only one VBS
~\cite{Suzuki:2015}.

The measured extinction ratio of 60.5 dB is at the limit of our temperature stability and our voltage drivers' resolution, and we are investigating ways to improve our measurement further. The sinusoidal nature of the phase control relationship works against us for achieving 50:50 splitting ratio VBS because this is at the point where the rate of change of splitting ratio with respect to voltage is the greatest. Improved voltage stability is required to overcome this. 
We are also investigating an improved resolution voltage driver, as it is in the deepest region of the fringe that a small change in the applied voltage can induce a sharp change in the optical power. There is thus plenty of scope for further enhancing the performance of our MZI.

\section*{Conclusion}
We have experimentally demonstrated a one-part-in-a-million $60$ dB high-extinction ratio MZI (in spite of imperfect fabrication of beamsplitters) using a self-adjustment approach and without calibrating any components (Fig \ref{fig:HighXFringe}). 
The result has many potential applications in several fields, from the realisation of high-quality photonics for telecommunication, to the development of quantum photonic technologies.The paper containing the original proposal of the beamsplitter 50:50 setup algorithm \cite{Miller:2015} also presents a further algorithm for optimising a larger optical circuit. The optimized VBS MZI such as ours can be cast as a functional building block which is repeated across a triangular array to form a Reck circuit for universal linear optics~\cite{Reck1984}. By virtue of the VBS MZI's improved extinction, one can access proportionately more of the total state space. Calibration of complex many-mode Reck schemes with high-extinction VBS MZIs is easier and more accurate. It is also likely to be quicker because of the reduced propagation of errors in measurements. These reasons illustrate how using the VBS MZI increases the maximum size of any meaningful and useful Reck scheme.

The realization of near-perfect beamsplitters and MZIs is essential to reduce the cumulative error in quantum operations, thus resulting in a significant reduction of overhead resources for fault-tolerant linear optical quantum computing~\cite{Raussendorf, Terry,Terry2}.  
This is because the number of error-correcting qubits scales inversely with the operation errors which are associated with the performance of beamsplitters \cite{Terry2}. Generally, this self-adjustable interferometric architecture offers potential to become a building-block component yielding the large-scale integration of linear-optical circuits for quantum information processing. 

\bigskip
\noindent * callum.wilkes@bristol.ac.uk

\noindent $\dagger$ jeremy.obrien@bristol.ac.uk

\begin{spacing}{0.}
\bibliographystyle{plain}

\bibliographystyle{unsrt}

\begin{thebibliography}{10}


\bibitem{Thomson:2016}
D.~Thomson, \emph{et al}.
\newblock{J. Opt.} 18, 073003, 2016.

\bibitem{Agrell:2016}
E.~Agrell, \emph{et al}.
\newblock{J. Opt.} 18, 063002, 2016.

\bibitem{Miller:2009}
D.~A.~B.~Miller.
\newblock{Proc. IEEE} 97, 1166 - 1185, 2009.

\bibitem{Krishnamoorthy:2015}
A.~V.~Krishnamoorthy \emph{et al}.
\newblock{J. Lightwave Technol.} 33, 889-900, 2015.


\bibitem{ITRS:2015}
International Technology Roadmap for Semiconductors 2.0 2015 Edition
\newblock{http://www.itrs2.net/itrs-reports.html}


\bibitem{bennett1984quantum}
C.~H.~Bennett,
\newblock {\em International Conference on Computer System and Signal Processing, IEEE, 1984}: 175--179, 1984.


\bibitem{korzh2015provably}
B.~Korzh \emph{et al}.
\newblock {\em Nature Photonics}, 9 (3): 163--168, 2015.

\bibitem{holland1993interferometric}
M.~J.~Holland
\newblock {\em Physical review letters}, 71 (9): 1355, 1993.

\bibitem{dowling2008quantum}
J.~P.~Dowling,
\newblock {\em Contemporary Physics}, 49 (2): 125--143, 2008.

\bibitem{knill2001scheme}
E. ~Knill, \emph{et al}.
\newblock {\em Nature}, 409 (6816): 46--52, 2001.

\bibitem{NielsenChuang}
M.~A. Nielsen and I.~L. Chuang.
\newblock{Quantum computation and quantum information} \newblock{\em Cambridge University Press}, 2010.

\bibitem{OBrienreview}
J. L. O'Brien, \emph{et al}.
\newblock {\em Nat. Photon.} 3: 687--695, 2009. 

\bibitem{Politi}
A. ~Politi, \emph{et al}. 
\newblock {\em Science}, 320: 646--649, 2008.

\bibitem{Shadbolt}
P.~J. Shadbolt, \emph{et al}.
\newblock {\em Nat. Photon.}, 6 (1): 45--49, 2012. 

\bibitem{Reck}
J. Carolan, \emph{et al}. 
\newblock {\em Science} 349: 711--716, 2015. 

 
\bibitem{Silverstone2014}
 J.~W. Silverstone, D.~Bonneau, \emph{et al}. 
\newblock {\em Nat. Photon.}, 8: 104--108, 2014. 

\bibitem{SiBell}
J.W. Silverstone, R. Santagati, \emph{et al}. 
\newblock {\em Nat. Commun.} 6: 7948, 2015.

\bibitem{Sharping}
J.~E. Sharping, \emph{et al}. 
\newblock {\em Opt. Express}, 14 (25): 12388--12393, 2006.


\bibitem{Tang}
W.~H.~P. Pernice, \emph{et al}. 
\newblock {\em Nat. Commun.}, 3 (1325): 1--10, 2012.

\bibitem{Dirk}
F. Najafi, \emph{et al}. 
\newblock {\em Nat. Commun.}, 6 (5873): 1--8, 2015.

\bibitem{Wang2015}
J. Wang, \emph{et al}. 
\newblock{\em Optica } 3: 407--413, 2016. 

\bibitem{QKD}
P. Sibson, \emph{et al}.
\newblock {arXiv}:1509.00768, 2015. 

\bibitem{Chen}
L. Chen and Y.-K. Chen. 
\newblock {\em Opt. Express} 20 (17): 18977--18985, 2012. 


\bibitem{Suzuki:2014}
K. Suzuki, \emph{et al}. 
\newblock {\em Opt. Express} 22 (4): 3887--3894, 2014. 


\bibitem{Raussendorf}
R.~Raussendorf and H.~J. Briegel.
\newblock {\em Phys. Rev. Let.}, 86 (22): 5188--5191, 2001.

\bibitem{Terry}
D. E. Browne and T. Rudolph. 
\newblock {\em Phys. Rev. Lett.} 95: 010501, 2005.


\bibitem{Terry2}
T. Rudolph. 
\newblock {arXiv}:1607.08535, 2016. 


\bibitem{Suzuki:2015}
K. Suzuki \emph{et al}.
\newblock  {\em Opt. Express}, 23 (7): 9086--9092, 2015.

\bibitem{Miller:2013}
D.~A.~B Miller.
\newblock {\em Photon. Res.} 1: 1--15, 2013.  

\bibitem{Miller:2015}
D.~A.~B Miller.
\newblock {\em Optica}, 2: 747--750, 2015.


\bibitem{Taillaert:2003}
D. Taillaert, \emph{et al}.
\newblock{IEEE Photon. Technol. Lett.}
15, 1249--1251 (2003).

\bibitem{QSim2016a}
S. ~Paesani, A. ~A. ~Gentile, \emph{et al}.
\newblock {\em In preparation}

\bibitem{Reck1984}
M.~Reck \emph{et al}.
\newblock {\em Phys. Rev. Lett.}, 73 (1): 58--61, 1994.


\end{thebibliography}

\end{spacing}

\section*{Acknowledgements}
This work was supported by the Engineering and Physical Sciences Research Council (EPSRC), a Marie Curie ITN in Photonic Integrated Compound Quantum Encoding (PICQUE), a Multidisciplinary University Research Initiative grant (Air Force Office of Scientific Research, FA9950-12-1-0024), the U.S. Army Research Office (ARO) grant W911NF-14-1-0133, and the Centre for Nanoscience and Quantum Information (NSQI). M.G.T acknowledges fellowship support from the Engineering and Physical Sciences Research Council (EPSRC, UK). J.L.O`B. acknowledges a Royal Society Wolfson Merit Award and a Royal Academy of Engineering Chair in Emerging Technologies.

\end{document}